\documentclass[conference,compsoc]{IEEEtran}
\usepackage[T1]{fontenc}
\usepackage{amsmath,amssymb}
\usepackage{booktabs,array,tabularx}
\usepackage{graphicx}
\usepackage{balance,lastpage}
\usepackage{listings}
\lstdefinestyle{tv}{basicstyle=\ttfamily\footnotesize\linespread{1.08}\selectfont,numbers=left,numberstyle=\scriptsize\color{gray},
  numbersep=4pt,frame=single,framerule=0.3pt,rulecolor=\color{gray!60},
  commentstyle=\color{gray}\itshape,keywordstyle=\bfseries,breaklines=true,
  xleftmargin=1.2em,aboveskip=6pt,belowskip=6pt,columns=fullflexible,keepspaces=true,breakatwhitespace=true}
\makeatletter
\let\tv@listingcaption\lst@makecaption
\long\def\lst@makecaption#1#2{%
  \noindent\parbox[b]{\linewidth}{\tv@listingcaption{#1}{#2}\par}%
  \par\nobreak\vspace{8pt}%
}
\makeatother
\usepackage{tikz}
\usetikzlibrary{arrows.meta,positioning,fit,calc}
\PassOptionsToPackage{hyphens}{url}
\usepackage[hidelinks]{hyperref}
\usepackage{url}
\makeatletter
\g@addto@macro\UrlBreaks{\do\a\do\b\do\c\do\d\do\e\do\f\do\g\do\h\do\i\do\j\do\k\do\l\do\m\do\n\do\o\do\p\do\q\do\r\do\s\do\t\do\u\do\v\do\w\do\x\do\y\do\z}
\makeatother
\newcolumntype{L}[1]{>{\raggedright\arraybackslash}p{#1}}
\newcolumntype{Y}{>{\raggedright\arraybackslash}X}
\newcommand{\argus}{TrafficVET}
\newcommand{\oc}{Open Controller}

\newcommand{\key}[1]{\texttt{#1}}
\definecolor{ink}{RGB}{37,55,75}
\definecolor{pale}{RGB}{235,242,248}
\definecolor{risk}{RGB}{149,50,42}
\tikzset{box/.style={draw=ink,rounded corners=1pt,fill=pale,align=center,font=\small,inner sep=5pt},arr/.style={-{Latex[length=2mm]},draw=ink,thick},riskarr/.style={-{Latex[length=2mm]},draw=risk,thick}}
\newcommand{\cw}[1]{\href{https://cwe.mitre.org/data/definitions/#1.html}{CWE-#1}}
\title{A Systematic Security Analysis of Traffic Signal Controllers}
\author{\IEEEauthorblockN{Lu Gao}
\IEEEauthorblockA{University of Houston}
\and
\IEEEauthorblockN{Yongxin Liu}
\IEEEauthorblockA{Embry-Riddle Aeronautical University}}
\AddToHook{shipout/after}{\ifnum\value{page}=\numexpr\getpagerefnumber{LastPage}-1\relax\balance\fi}
\begin{document}
\maketitle
\begin{abstract}
Traffic signal controllers translate detector observations, engineering parameters, and supervisory commands into signal indications governing vehicle and pedestrian movements. We conduct a systematic security analysis of three traffic signal controllers: Open Controller, developed with the City of Helsinki and field-tested in its J{\"a}tk{\"a}saari living lab; SCOPE, a USDOT-funded dual-controller system field-tested at an Orlando intersection on ATC cabinet hardware; and openTLC. We identify 17 security-relevant software findings. We characterize the findings by CWE-mapped root causes, triggering conditions, and observed effects, and separately classify attacker access and validation scope. In the evaluated configurations, four findings are reachable through unauthenticated network inputs and four through detector-publisher inputs; one involves local engineering input examined in an isolated test. Seven additional findings arise during ordinary operation or faults; one has no established attacker-controlled entry. The findings span authorization failures, input validation, safety constraint enforcement, state and lifecycle management, and memory safety. Their demonstrated effects include conflicting green indications, insufficient vehicle clearance intervals, interrupted controller operation, persistent phase starvation, incorrect redundancy validation, and shortened pedestrian timing. We analyze the underlying software mechanisms and validate representative findings through original-controller execution and corrective-patch experiments. We developed TrafficVET, a testing framework that extracts controller constraints, drives the original software, and checks traces with independent safety oracles; its blind campaigns first surfaced three of the reported findings, including the pedestrian-clearance and restart defects. 
\end{abstract}
\section{Introduction}
Traffic signal controllers implement a continuous transfer of right of way. Every detector request, phase decision, and timing update ultimately contributes to an output sequence that road users interpret as permission to move. Safe operation relies on both spatial separation and temporal separation: incompatible movements must be kept apart, terminating vehicle movements require appropriate transitions, and pedestrians require time to complete a crossing. These properties make controller software a consequential boundary between networked infrastructure and physical traffic operations~\cite{fhwa,yellow,mutcd}.

We examine three controller implementations, two of which have been tested with live signal infrastructure. Open Controller was developed in the Smart Junction project by the City of Helsinki, Aalto University, and Conveqs, and has been used for field signal-control trials in the J{\"a}tk{\"a}saari living lab~\cite{smartjunction,conveqsjunction,conveqstrials2025}; its repository states that the interface to physical signal controllers is withheld for safety reasons~\cite{opencontroller}. SCOPE was developed under USDOT funding and field-tested at a signalized intersection in Orlando, Florida, on PEEK ATC-1000 controller hardware, as documented in the developer's 2013 report~\cite{scopefield}. The security of reusable operational-technology software is also a focus of current guidance~\cite{icsdefense,nist218}. \oc{} exposes a message-driven control engine; openTLC combines detector requests with configurable movement relationships; and SCOPE implements primary and secondary controllers in different programming languages~\cite{opencontroller,opentlc,scope}. Their architectures provide three distinct opportunities to examine how external inputs reach safety-critical state. For each controller, we trace input processing, runtime state, and output generation, and validate corrective changes.

Prior traffic-security research establishes that network access and falsified observations can materially affect signal operation. Ghena et al. demonstrated control of a deployed traffic system through weaknesses in its networked infrastructure~\cite{ghena2014}. Chen et al. showed how falsified connected-vehicle data changes signal scheduling and congestion~\cite{chen2018}. More recent work studies adversarial observations against learning-based signal control~\cite{li2025}. We investigate how accepted network, detector, and engineering inputs lead to violations of signal-transition, clearance, and controller-state rules.

Answering this question requires following an input beyond its acceptance by a protocol handler. An ordinary restart command can reset a live phase. A well-formed engineering message can carry a timing value that removes a protective interval. A controller can continue producing syntactically valid output while that output violates a safety requirement. We trace input handling, controller-state changes, emitted signal states, and the effects of corrective patches.

We systematically investigate three traffic signal controllers and characterize 17 security-relevant software findings, V01--V17. We organize the findings into five root-cause families and distinguish each finding's root cause, trigger, observed effect, attacker-access class, and validation scope.  We conduct detailed original-versus-corrected experiments for V03, V04, and V13.

We make two contributions:
\begin{itemize}
\item \textbf{Systematic security analysis.} We identify and validate 17 CWE-mapped software-security findings across three traffic signal controllers, covering authorization, input validation, safety-rule enforcement, controller state, and memory safety.
\item \textbf{TrafficVET: a constraint-grounded testing framework.} We developed TrafficVET, which generates tests from extracted controller constraints and checks execution traces using independent safety oracles. Its blind campaigns first detected V04, V05, and V13.
\end{itemize}

\section{System Model and Safety Properties}
\label{sec:background}
\subsection{Traffic Signal Operation}
A \emph{signal group} represents indications assigned to a movement or a set of movements, while a \emph{phase} specifies a service decision over compatible movements. Actuated control changes service in response to detector and pedestrian requests. Minimum-green and extension rules govern the current service interval; change and clearance intervals govern the transfer to another movement. Vehicle yellow, red clearance, pedestrian walk, and pedestrian clearance each express a different part of that transfer~\cite{fhwa}.

We represent an observed controller execution as a sequence of timestamped states. For group $g$, let $q_g(t)$ denote its indication and let $r_g(t)$ denote a pending request. Let $C_{ij}$ identify incompatible movements and let $D_{ij}$ denote the required interval between the end of green for movement $i$ and the next start of green for movement $j$. $C$ describes a relationship between movements; $D$ describes an ordered transition. The distinction is operationally important because travel distances and clearance requirements can differ by direction.

The safety interpretation of a parameter comes from an engineering profile and its corresponding expected behavior. In our SCOPE reproduction profile, the nominal vehicle yellow and red-clearance settings are each 5 seconds. The pedestrian profile uses a 5-second walk value and a 5-second clearance value, giving an expected combined interval of 10 seconds. 

\subsection{Traffic-Control Subsystems}
The traffic-control and detector handbooks describe the operational roles of local controllers, supervisory communications, and detector subsystems~\cite{signalhandbook,detectorhandbook}. Detector inputs, operator commands, and engineering settings enter the controller through different interfaces. The controller maintains phase state and emits signal indications.

\subsubsection{Subsystem Taxonomy}
The source code of all three controllers is publicly available.

\textbf{Open Controller.} \oc{}, developed in the Smart Junction project (City of Helsinki, Aalto University, Conveqs), implements a Python control engine, Clockwork, with phase-ring and signal-group state. The project includes a demonstration model of intersection 270 of the J{\"a}tk{\"a}saari test area~\cite{opencontroller}, and the project reports field signal-control trials in 2024--2025~\cite{smartjunction,conveqsjunction,conveqstrials2025}; the component that connects Clockwork to physical signal controllers is not published, for safety reasons~\cite{opencontroller}. Subsequent research has built on the software for signal-parameter optimization~\cite{lattu2025}. NATS subjects carry detector data, lifecycle commands, and group-control outputs. The project supports integration with simulation and controller interfaces~\cite{opencontroller}. For our analysis, this architecture permits observation at both the controller-state boundary and the message-consumer boundary. The real-mode test path starts original Clockwork and a NATS broker, delivers messages, and records emitted control states.

\textbf{openTLC.} openTLC is a Python traffic-control project whose engineering configuration contains movement identifiers, conflict relationships, and directional clearance values~\cite{opentlc}. Its detector-to-group mapping determines how an input becomes a service request. SUMO network geometry supplies an additional representation of the intersection, allowing selected movement relationships to be evaluated independently of the controller's configured conflict table. SUMO provides network and traffic simulation facilities relevant to this comparison~\cite{sumo}.

\textbf{SCOPE.} SCOPE, developed by Advanced Technologies Inc. under USDOT funding, combines an Ada primary controller and a C++ secondary controller. The developer's 2013 final report documents field testing at the Oak Ridge Road / Harcourt Avenue intersection in Orlando, Florida, on PEEK ATC-1000 controller hardware~\cite{scopefield}. The project provides actuated and fixed-time operation, simulated detector and pedestrian calls, and graphical engineering interfaces~\cite{scope,scopeprogram}. Our study uses SCOPE 1.1 with its original 32-bit primary/secondary build. A GUI protocol delivers settings and calls to the primary, and timestamped traces record primary and secondary state.

\subsection{Controller Network Architecture}
Open Controller uses subject-based messaging, with a text client protocol and a payload carried between publishers and subscribers~\cite{natssubject,natsprotocol}. SCOPE delivers its engineering record over TCP, whose reliable byte-stream semantics place record-boundary handling in the application~\cite{rfc9293}. Input serialization and field interpretation are therefore examined separately; JSON and YAML specify data representations used by the surrounding controller and experiment configurations~\cite{rfc8259,yaml}. Three property families guide the investigation. \emph{Mutual exclusion} concerns simultaneous indications for incompatible movements. \emph{Transition timing} concerns yellow, intergreen, and pedestrian intervals. \emph{Progress and continuity} concern sustained requests, controller liveness, and state preservation across lifecycle operations. Runtime availability complements these properties by identifying abnormal termination of the process responsible for producing indications.

These families require different observations. A conflict check uses movement identities and colors. A clearance check uses an ordered pair and the timestamps at which its two relevant transitions occur. A recovery check needs the state immediately before and after a command. A liveness diagnosis combines request history with the controller's blocked state.

SUMO represents road networks and signal control explicitly~\cite{sumonet,sumotl}. Its TraCI interface exposes simulation interaction to external clients~\cite{traci}. Signal-performance guidance likewise connects operational assessment to recorded signal behavior~\cite{atspm,signalsafety}.

\section{Security Analysis Framework}
\label{sec:position}
\subsection{Systematization of Traffic-System Security Research}
Traffic-system security research examines network access, falsified observations, and control decisions. Green Lights Forever demonstrates control of signal infrastructure through network and device weaknesses~\cite{ghena2014}. Connected-vehicle studies analyze falsified trajectories~\cite{chen2018}, and adversarial-learning studies examine cooperative-perception inputs and adaptive controllers~\cite{li2025}. Other attacks use surrogate models~\cite{huang2021}, slow-poisoning detector inputs~\cite{dasgupta2021}, and Sybil-vehicle injection~\cite{irfan2022}. Signal-tampering optimization measures network throughput~\cite{thodi2020}; security games and resilient control examine detection and mitigation~\cite{securitygame,resilience}. Laser illumination attacks target traffic-light recognition at the receiving vehicle~\cite{laser}. Our analysis examines how accepted inputs propagate through controller code to signal outputs, service decisions, and redundant-state validation.

\subsection{Controller Input and State Analysis}
Control-aware testing checks failures that preserve ordinary program execution. RVFuzzer searches for parameter-validation defects using control-guided mutation~\cite{rvfuzzer}; PGFUZZ encodes safety and functional policies as temporal constraints~\cite{pgfuzz}. AFLGo directs execution toward selected program locations~\cite{aflgo}. SAVIOR monitors physical invariants, and DriveFuzz evaluates driving behavior~\cite{savior,drivefuzz}. ICSREF and ICSFuzz analyze and test industrial control binaries~\cite{icsref,icsfuzz}; FIRMADYNE, FIRM-AFL, P2IM, and Fuzzware provide hardware-independent execution and firmware-testing methods~\cite{firmadyne,firmafl,p2im,fuzzware}.

Stateful testing retains input sequences and execution history, as in AFLNet, StateAFL, and stateful greybox fuzzing~\cite{aflnet,stateafl,sgfuzz}. AFL++ and libFuzzer explore software inputs~\cite{aflpp,libfuzzer}; AddressSanitizer detects invalid memory accesses~\cite{asanpaper,asan}. We examine controller input handlers, configuration-to-state assignments, phase transitions, and primary-secondary comparisons. Controller traces supply indication order and timing; component tests exercise validation functions and callback lifetimes.

\subsection{Three-Dimensional Finding Characterization}
We characterize each finding along three analytical dimensions. \emph{Root Cause} identifies the faulty software mechanism and its CWE mapping. \emph{Triggering Condition} identifies the input, event, or operating state that activates the mechanism, separately from any privilege needed to supply that input. \emph{Observed Effect} records the resulting change in signal timing, movement compatibility, controller availability, redundancy validation, or memory integrity. We also record \emph{Attacker Access} and \emph{Validation Scope} as separate attributes, distinguishing externally triggered failures from normal-operation defects and isolated component-test results.

Findings are counted by distinct software failure mechanisms. Multiple triggering inputs or affected signal groups associated with the same mechanism are represented as one finding. We organize the 17 core findings into five root-cause families: authorization, input validation, safety-rule enforcement, state and lifecycle management, and memory safety.

The mappings follow the code operation: missing command checks use CWE-862~\cite{cwe862}; quantity validation uses \cw{1284}; wrong comparison factors use \cw{1025}; synchronization uses \cw{821}; failure propagation uses \cw{636}; and the memory findings use \cw{787} and \cw{825}. CWE-841 provides the workflow and timing category for signal-state enforcement~\cite{cwe841}. Input validation, exceptional-condition handling, authentication, and calculation weaknesses supply related categories for examining the receiving boundary~\cite{cwe20,cwe754,cwe306,cwe682}.

\section{Finding Taxonomy and Triggering Surfaces}
\label{sec:threat}
\subsection{Root-Cause Families}
The inventory contains one authorization finding, three input-validation findings, five safety-rule enforcement findings, six state/lifecycle findings, and two memory-safety findings. Open Controller contributes eight findings, openTLC two, and SCOPE seven. Authorization controls who can deliver a command. Input checks determine which values enter runtime state. Timing and conflict rules constrain output transitions; lifecycle and fail-safe logic govern continuity and recovery; object and array operations govern execution integrity.

\subsection{Controller Safety Boundaries}
Open Controller accepts detector observations and supervisory commands through NATS, updates phase-ring and signal-group objects, and publishes group-control states. openTLC reads local timing and conflict configuration and converts SUMO or GPIO detector calls into phase decisions. SCOPE receives engineering records over TCP and compares Ada-primary state with C++-secondary state. Its callbacks and detector-to-phase assignments connect protocol values to the secondary's memory objects.

\begin{figure*}[!t]
  \centering
  \includegraphics[width=4.15in]{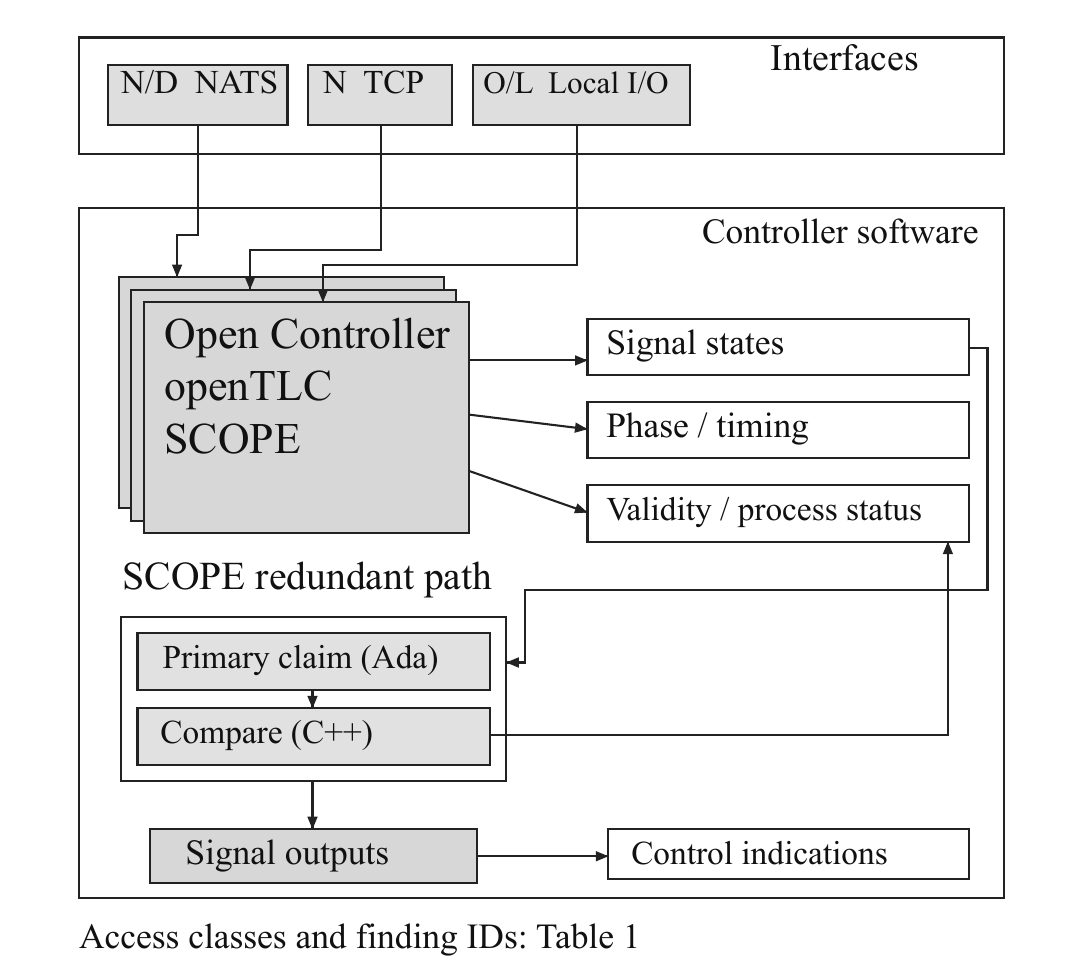}
  \caption{Controller interfaces and software boundaries in the three studied targets. Open Controller receives NATS detector and lifecycle messages; SCOPE receives TCP engineering records; openTLC uses local detector calls and configuration. The nested SCOPE path compares primary state claims in the C++ secondary. Signal indications, phase timing, validity, and process status provide observations for testing; Table~\ref{tab:findings} gives each finding's access class.}
  \label{fig:surfaces}

\end{figure*}
\subsection{Threat Model and Access Classification}
Figure~\ref{fig:surfaces} maps controller interfaces to the access classes in Table~\ref{tab:findings}.
We classify the minimum access used to trigger each finding in the evaluated setup. \textbf{N (unauthenticated network)} requires network reachability but no credentials. \textbf{D (detector publisher)} requires permission to publish detector messages, without lifecycle-command or output-control privileges. \textbf{L (local engineering)} requires control over local engineering inputs or configuration. \textbf{O (ordinary operation or fault)} applies when the behavior arises without adversarial input. \textbf{U (unestablished attacker entry)} denotes a component-level defect for which no attacker-controlled entry has been demonstrated. These labels describe access rather than severity.

\textbf{Evaluated Input Interfaces.} We evaluate Open Controller's NATS message subjects, SCOPE's TCP engineering interface, and openTLC's local detector-call path. Open Controller is designed to separate data processing from signal control, with NATS carrying detector data, lifecycle commands, and group-control outputs between components; in the J{\"a}tk{\"a}saari trials this message boundary sits between the control engine and the (unpublished) adapter to the physical controller~\cite{opencontroller}, so NATS subject reachability is the field attack surface for V01 and V13. SCOPE's TCP engineering interface on port 5878 is the path through which its GUI delivers settings and calls to the primary~\cite{scope}; the Orlando test used the primary/secondary system on ATC hardware~\cite{scopefield}. We execute the original controller software in containers and local test processes. Cabinet-level MMU/CMU protection is outside the evaluated software path. SCOPE's secondary controller is its software analogue of that monitor, and V14, V15, and V09 show that this analogue does not validate at startup, does not force a safe state on secondary loss, and compares the wrong ring field.

In the tested Open Controller configuration without NATS access controls, an input publisher can reach lifecycle and control subjects; V01 and V13 exercise this boundary. Restricted detector publishers can supply the inputs relevant to V02, V05, V11, and V12, without permission to publish privileged command or output subjects~\cite{nats}. The SCOPE primary listens on all interfaces (0.0.0.0) on TCP port 5878 and accepts unauthenticated 777-byte engineering records; V03 and V04 are reproduced through this interface. V07 and V08 are triggered by detector requests during local openTLC controller execution.

Table~\ref{tab:findings} records both Attacker Access and Triggering Condition. For findings with both attacker-controlled and ordinary triggers, the access code describes the demonstrated attacker-controlled path. We additionally mark validation scope: \textbf{E} denotes execution or replay of original controller software in the test environment~\cite{docker}; \textbf{C} denotes an isolated compiled component or sanitizer test. V09, V16, and V17 have C-level evidence. V16 uses a local engineering-input path and is validated through an isolated AddressSanitizer fixture. The V15 process-loss experiment uses a controlled SIGKILL event to test the primary controller's fault response. Across the 17 findings, the access classes are N=4, D=4, L=1, O=7, and U=1; validation scopes are E=14 and C=3.

\subsection{Representative Original-versus-Corrected Cases}
\begin{figure*}[!t]
  \centering
  \includegraphics[width=\textwidth]{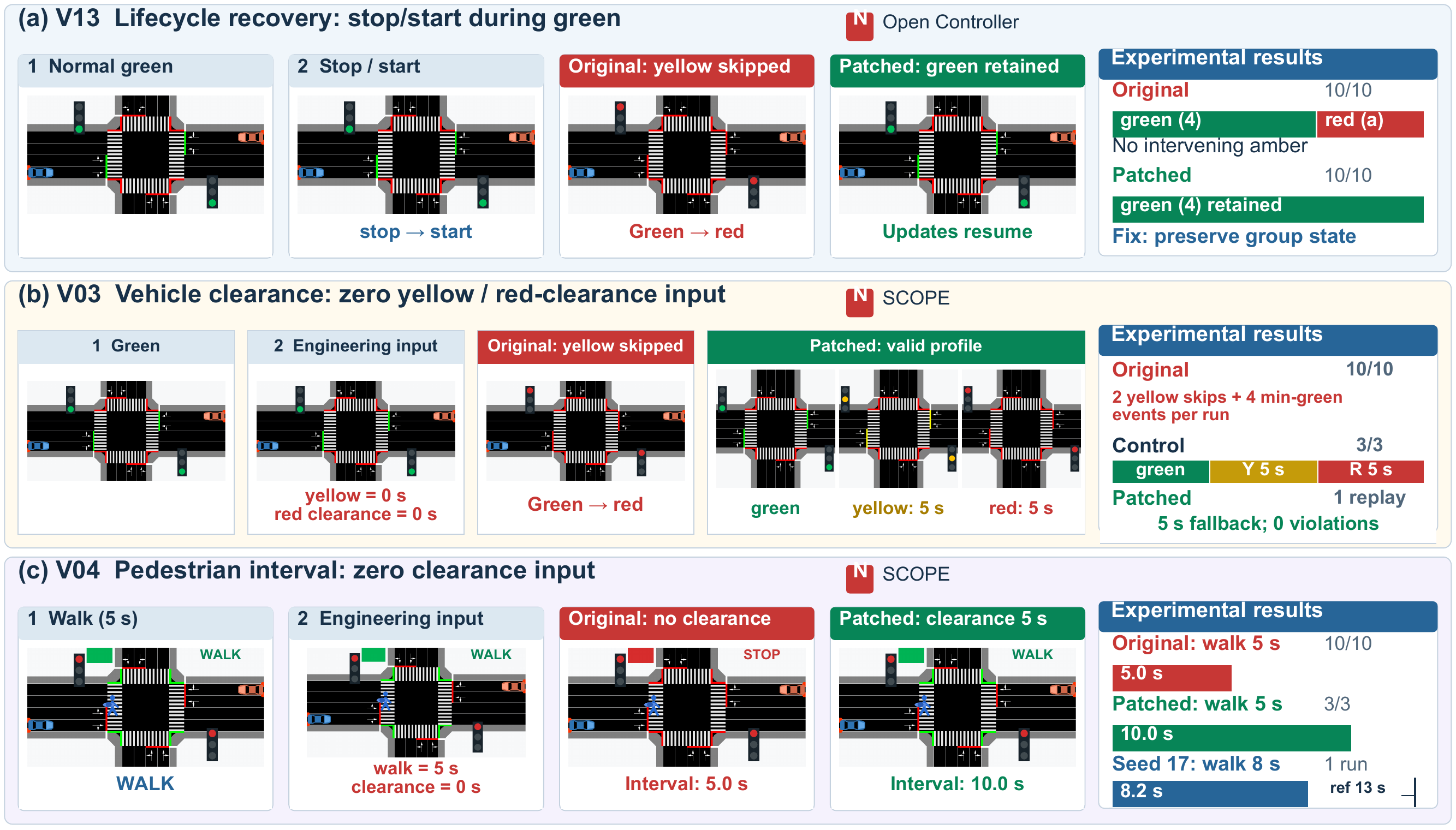}
  \caption{Traffic-signal storyboards and recorded results for (a) V13, unsafe restart recovery; (b) V03, zero vehicle clearance; and (c) V04, zero pedestrian clearance. Road scenes are schematic. V13 reproduces in 10/10 original runs and retains green in 10/10 patched runs. V03 uses 10 original runs, three control runs, and one patched replay. V04's targeted trace-reported intervals are 5.0 versus 10.0 seconds; blind seed~17 records an 8.2-second active interval against a 13.0-second reference.}
  \label{fig:representative}
\end{figure*}

Figure~\ref{fig:representative} illustrates three representative security failures: unsafe restart recovery (V13), vehicle-clearance parameter validation (V03), and pedestrian-clearance parameter validation (V04). Table~\ref{tab:findings} presents the 17-finding core inventory. The restart correction retains live green state; the vehicle-clearance correction restores yellow; the pedestrian correction restores the trace-reported combined duration from 5.0 to 10.0 seconds.

\section{Security Findings Across Controllers}
\label{sec:assessment}\label{sec:impact}
\subsection{Finding Inventory and Classification}
Table~\ref{tab:findings} lists the 17 core findings in root-cause order. For each finding, the table reports its CWE-mapped root cause, attacker-access class, triggering condition, observed effect, validation scope, and supporting test. Access classes separate adversarial input capabilities from routine controller events, while validation scope distinguishes controller-level execution from isolated component tests. The analysis covers message-driven Python control, detector-driven Python control, and Ada/C++ redundant control~\cite{opencontroller,opentlc,scope,scopeprogram}. For each controller, we analyze executable phase logic, input pathways, and observed state transitions.
\begin{table*}[!t]
\caption{Security Findings Across Three Traffic Signal Controllers}
\label{tab:findings}
\centering\footnotesize
\renewcommand{\arraystretch}{1.10}
\setlength{\tabcolsep}{2pt}
\begin{tabularx}{\textwidth}{@{}L{6mm}L{12mm}L{34mm}L{9mm}L{27mm}Y L{9mm}L{22mm}@{}}
\toprule
ID & Controller & Root Cause / CWE & Access & Trigger Condition & Observed Effect & Scope & Validation \\
\midrule
\multicolumn{8}{@{}l}{\textbf{Authorization}} \\
V01 & OC & Missing NATS Command Authorization (\cw{862}) & N & Unauthorized NATS command/control publication & Unauthorized stop; external control of signal indications & E & Original execution; ACL replay \\

\addlinespace[2pt]
\multicolumn{8}{@{}l}{\textbf{Input validation}} \\
V02 & OC & Invalid Detector Count Handling (\cw{1284}) & D & Invalid detector-count payload & Abnormal controller termination & E & Original/fixed input replay \\
V03 & SCOPE & Vehicle Clearance Parameter Validation Failure (\cw{1284}) & N & Zero vehicle yellow/red-clearance values & Yellow skipped; vehicle timing violations & E & 10 original runs; fixed replay \\
V04 & SCOPE & Pedestrian Clearance Parameter Validation Failure (\cw{1284}) & N & Zero pedestrian-clearance value & Pedestrian duration field: 10 s to 5 s & E & 10 original / 3 fixed runs \\

\addlinespace[2pt]
\multicolumn{8}{@{}l}{\textbf{Safety-rule enforcement}} \\
V05 & OC & Directional Intergreen Indexing Error (\cw{841}) & D & Detector calls under asymmetric clearance & Insufficient directional clearance & E & Original/fixed controller replay \\
V06 & OC & Incorrect Amber-Red Timing Reference (\cw{841}) & O & Normal phase transition & 7.1 s clearance against 8 s requirement & E & Timing replay; combined fix \\
V07 & openTLC & Directional Clearance Enforcement Failure (\cw{841}) & O & Vehicle detector requests & 54 intergreen violations over 19 pairs & E & Original/fixed 20,000-step run \\
V08 & openTLC & Missing Geometric Conflict Constraints (\cw{841}) & O & Concurrent crossing-movement requests & Overlapping greens for three crossing pairs & E & Three-pair original/fixed replay \\
V09 & SCOPE & Incorrect Secondary Ring Color Comparison (\cw{1025}) & U & Ring-2 mismatch in compiled validator test & Validator accepts inconsistent ring state & C & Compiled original/fixed unit test \\

\addlinespace[2pt]
\multicolumn{8}{@{}l}{\textbf{State and lifecycle management}} \\
V10 & OC & Unsafe Startup Phase Transition (\cw{841}) & O & Controller startup & Rapid phase changes; conflicting greens & E & Original/fixed startup replay \\
V11 & OC & Persistent-Demand Phase Starvation (\cw{841}) & D & Sustained detector demand & Conflicting phase unserved for 600 s & E & Targeted original/fixed replay \\
V12 & OC & Stale Next-Phase State Lock (\cw{841}) & D & Transient request retaining next-phase state & Persistent controller lock; unserved demand & E & Original/fixed state trace \\
V13 & OC & Unsafe Restart State Reset (\cw{841}) & N & NATS stop/start during active green & Green directly to red; amber omitted & E & 10 original / 10 fixed runs \\
V14 & SCOPE & Primary-Secondary Synchronization Failure (\cw{821}) & O & Startup with unsynchronized controller states & Primary/secondary phase and color disagreement & E & Dual-controller traces; fault injection \\
V15 & SCOPE & Secondary Failure Propagation Failure (\cw{636}) & O & Secondary process loss (SIGKILL in test) & Primary stays green with validity set & E & Original/fixed process-loss test \\

\addlinespace[2pt]
\multicolumn{8}{@{}l}{\textbf{Memory safety}} \\
V16 & SCOPE & Out-of-Bounds Pedestrian Phase Assignment (\cw{787}) & L & Out-of-range engineering phase assignment & Secondary out-of-bounds write and crash & C & ASan isolated test; bounds fix \\
V17 & SCOPE & Dangling Controller Callback (\cw{825}) & O & Normal status-request callback & Secondary accesses an expired controller object & C & ASan status test; lifetime fix \\
\bottomrule
\end{tabularx}
\smallskip
\parbox{\textwidth}{\footnotesize OC: Open Controller. Access: N, unauthenticated network; D, detector publisher; L, local engineering; O, ordinary operation/fault; U, no established attacker entry. Scope: E, controller-level execution/replay; C, isolated component/sanitizer test. V14's corrective work remains partial.}
\end{table*}

\subsection{Finding Analysis and Validation}
\label{sec:analysis}
 Detailed original-versus-corrected experiments are presented for V03, V04, and V13, together with additional controller-specific validation results.

For each finding, we trace the input or operating condition through the receiving handler, runtime assignment, and affected state transition. The validation records controller traces, process exit, comparison results, or sanitizer output. Paired experiments apply the same relevant input to original and corrected code. V01 groups command and output-subject effects under one missing-authorization mechanism. V03 and V04 share a transport but modify separate vehicle and pedestrian timing variables. V11 affects service continuity through blocked phase progression under persistent demand.

\subsection{Open Controller Findings}
\label{sec:recovery}

\subsubsection{V05: Directional Intergreen Indexing Error}\label{sec:v05}
\textbf{Discovery.} The Open Controller main/fast campaign issues detector calls against the extracted intergreen matrix. Across 20 seeds the intergreen oracle recorded 13 short-clearance events; the evaluator's directional predicate, which requires a positive configured clearance in one direction and zero in the reverse, matched six of them. At seed 13, group 2 began green 0.0 s after group 6 ended green against a 1.0-second requirement, and the two groups held green together for four samples; seeds 1 and 10 recorded the same pair family at 0.2 s. A symmetric reading of the matrix would not flag these pairs. Source inspection, prompted by these events, then showed the lookup indexing the matrix in the wrong order.

V05 concerns the use of directional intergreen values when establishing the movement relation enforced during execution. An ordered clearance can be positive in one direction and zero in the reverse direction. The assessment therefore retains both the direction-specific requirement and the executed movement pair. Detector calls exercise the configured relation, and the root predicate associates the observed anomaly with the asymmetric configuration.

\subsubsection{V01: Missing NATS Command Authorization}\label{sec:v01}
V01 identifies the shared authorization family affecting NATS command and control subjects. In the tested no-ACL configuration, an input publisher can deliver control messages, including a command that stops the controller. A separate authorization probe checks the same subjects under restricted permissions and verifies that the restricted account's publications are rejected while the controller account retains its intended capability.

\subsubsection{V02: Invalid Detector Count Handling}\label{sec:v02}
V02 follows invalid detector data into integer conversion. A payload that reaches this conversion can terminate the original execution, and the generic campaign records four matching exits.  

\subsubsection{V11: Persistent-Demand Phase Starvation}\label{sec:v11}
V11 combines persistent demand with an extending green, delaying a conflicting phase's service.  In the targeted run, the competing phase remains unserved for 600 seconds; Section~\ref{sec:evaluation} reports the resulting generic-campaign outcome.

\subsubsection{V12: Stale Next-Phase State Lock}\label{sec:v12}
V12 concerns a next-phase index retained after a one-tick request. The real extender execution can remain locked with its current phase equal to that retained next phase. The oracle flags persistent locking when the current phase equals the retained next-phase index.

\subsubsection{V13: Unsafe Restart State Reset}\label{sec:v13}
\textbf{Discovery.} The extender/real campaign draws stop and start from its generic lifecycle action set. Four seeds (7, 11, 17, 19) selected that family; in three of them a stop arrived during an active green and the subsequent start produced a green-to-red change with no intervening amber in the emitted group-control messages---group B at 6.97 s (seed 7, with a 1.42-second green against a 2-second minimum), group A at 10.78 s (seed 11), and group A at 10.69 s (seed 17). The oracle detected three green-to-red transitions; source inspection traced them to signal-group reinitialization in the restart handler.

 V13 is reachable through the \key{clockwork.command} lifecycle interface in the tested no-ACL configuration. A stop command suspends updating while a group is green. The subsequent start command resets the timer and calls the controller's reset routine. That routine reconstructs the phase-ring controller and its groups, initializing their indications to red. The current green is consequently discarded during recovery.

The original sequence for group A is recorded as
\begin{equation}
 a\rightarrow b\rightarrow c\rightarrow f\rightarrow g
 \rightarrow 0\rightarrow 1\rightarrow 5\rightarrow 4\rightarrow a
\end{equation}
where the relevant final transition is $4\rightarrow a$: green directly to red. The emitted sequence contains no intervening amber state, encoded as \key{<}. These symbols are implementation states; the interpretation of the final transition comes from the controller's output mapping. Figure~\ref{fig:representative} presents the command and output sequence responsible for that change.

\textbf{Reproduction and correction.} The direct transition occurs in all ten original-software runs using original Clockwork, NATS, and an output receiver. The minimal correction resumes updating without calling the group-reinitialization path. All ten fixed traces end at green substate \key{4}, with the sequence $a\rightarrow b\rightarrow c\rightarrow f\rightarrow g\rightarrow 0\rightarrow 1\rightarrow 5\rightarrow 4$. The correction retains the existing group state while resuming updates. The normal baseline contains 52 control messages with zero oracle violations.

The restart-state defect also affects authorized stop/start commands because command authorization does not change the reset handler. Listing~\ref{lst:v13} shows the original start branch. 

\begin{lstlisting}[style=tv,language=Python,
  caption={Open Controller \texttt{clockwork.py}: the \texttt{start} handler re-runs
  \texttt{PhaseRingController.\_\_init\_\_}, which recreates every signal group in
  red (V13). The patch keeps the timer reset and resumes updates without the loop.},
  label={lst:v13}]
elif command == "start":
    print("Starting the controller")
    self._timer.reset()
    for controller in self._controllers:
        controller.reset()
    self._update = True
    self._state_changed.set()
\end{lstlisting}
\subsubsection{Timing and Startup-State Enforcement}
Two additional Open Controller findings concern timing and startup-state enforcement. V06 starts the amber-red timing reference before the actual green termination, producing a 7.1-second clearance against an 8-second requirement. V10 permits rapid startup phase transitions and conflicting green indications. V06 arises from using an earlier substate transition as the timing reference. V10 arises from an incomplete startup-state check; the original integrated run records 17 phase changes within eight seconds. 

\subsection{openTLC Findings}
\label{sec:opentlc}
\subsubsection{Directional Clearance and Geometric Conflicts}
\label{sec:geometry}
The openTLC investigation identifies two core findings. V07 concerns the omission of configured directional clearance times from executable phase-control logic. Original-controller execution records 54 intergreen violations across 19 movement pairs; the corrected controller records zero violations. V08 concerns missing geometric conflict relationships in the configured conflict matrix. Three confirmed crossing-movement pairs exhibit 487, 626, and 487 overlapping-green steps, respectively, all reduced to zero after correcting the conflict map.

For V07, phase logic closes yellow using the base yellow timer but does not enforce the configured ordered clearance before the next green. The 20,000-step baseline includes a 31-step interval from \key{fc32} to \key{fc01} against a 70-step requirement. V08 uses the crossing pairs \key{fc02/fc37}, \key{fc05/fc11}, and \key{fc11/fc35}; both directions are requested during the first 1,000 steps of each 1,200-step replay. The corrected conflict map eliminates overlaps while preserving service to both movements.

Repeated V07 replay in the same process clears historical green-end timestamps at step zero and records zero violations in both runs.

\subsection{SCOPE Findings}
\subsubsection{V03: Vehicle Clearance Parameter Validation Failure}
\label{sec:v03}
\textbf{Entry and propagation.} V03 uses the original GUI service. The primary accepts a 777-byte record containing operational commands, calls, and timing fields. In the tested path, a remote sender supplies zero values for the vehicle yellow and red-clearance fields. The primary assigns these values to the runtime timing variables used during phase execution. The attack therefore follows a normal engineering-input path into the state machine.

\textbf{Observed effect.} In the original trace, a pair of active green phases is followed by a red state without an intervening yellow sample. Across ten repeated runs, the oracle records two yellow-skip events and four minimum-green violations per run. The corresponding conflict-green count is zero. 

\textbf{Control and correction.} Three control runs using 5-second yellow and red-clearance settings record zero violations for these checks. A minimal correction validates both initialization-time and repeated GUI updates, replacing non-positive values with the 5-second test-profile fallback. Rebuilding and replaying the same zero-clearance input restores the yellow transition and produces zero recorded violations in the fixed replay. A container-to-container reproduction verifies delivery through the network-facing service.

\subsubsection{V04: Pedestrian Clearance Parameter Validation Failure}
\label{sec:v04}
\textbf{Discovery.} The SCOPE/real campaign samples pedestrian clearance from $\{0,1,2,5,8,10\}$ and walk from $\{3,5,8,10\}$, treating zero as an ordinary boundary value. Seeds 4, 8, 11, and 17 drew a zero clearance. At seed 17 the record carried an 8-second walk, a zero clearance, and a pedestrian call on phase 2; the pedestrian-interval oracle flagged an 8.2-second active interval against a 13.0-second reference (walk plus the profile's 5-second minimum clearance). The other three zero-clearance seeds placed the pedestrian call on phase 1, which the 20-second horizon did not reach.  Source analysis then traced the accepted value to the unvalidated per-phase assignment described below, and the targeted experiment reproduced the mechanism with a 5-second walk in ten of ten runs.

\textbf{Entry and propagation.} V04 uses the GUI service to set the pedestrian clearance value to zero while retaining a 5-second walk value. Vehicle and pedestrian calls activate the affected schedule. The source computes the total pedestrian interval as
\begin{equation}
 T_{\mathrm{ped}}=T_{\mathrm{walk}}+T_{\mathrm{clear}}
\label{eq:ped}
\end{equation}
so accepting a zero clearance directly removes its contribution. The reproduction drives the original primary in a container from a separate host process.

\textbf{Observed effect.} All ten original runs record a 5.0-second pedestrian interval against the profile's expected 10.0 seconds. The five-second reduction equals the entire configured clearance component and halves the combined interval. The measured value is the combined pedestrian-duration field in the runtime trace; Equation~(\ref{eq:ped}) shows how zero clearance reduces it to the walk duration.

\textbf{Correction.} The corrected patch checks pedestrian walk and clearance values for every phase and substitutes the normal 5-second profile value for non-positive inputs. All three recorded fixed runs restore the 10.0-second interval. 

\begin{lstlisting}[style=tv,language=Ada,
  caption={SCOPE primary: the pedestrian interval is the sum of the two
  GUI-supplied fields; a zero clearance is assigned without validation (V04).},
  label={lst:v04}]
Total_Ped_Time := SDP.Ped_Clearance_Time(Ped_Id) +
                  SDP.Ped_Walk_Time(Ped_Id);
\end{lstlisting}
Listing~\ref{lst:v04} shows the original pedestrian-duration calculation.
\subsubsection{Redundant-State Validation}
V09 identifies an incorrect ring-color comparison in SCOPE's redundant-state validator. In a compiled test, primary colors (red, green) and secondary colors (red, red) are incorrectly accepted by the original validator. After the second-ring comparison is corrected, the same mismatch is rejected. 

\subsubsection{Synchronization and Fail-Safe State}
V14 concerns primary-secondary startup synchronization and produces repeated phase and color disagreements in original-controller traces. The original actuated run shows disagreements in all 83 timestamp-matched samples; the fixed-time run shows 55 disagreements among 86 samples. The original traces show persistent primary-secondary startup disagreement. Section~8.4 reports the correction evaluated under injected inconsistent-state conditions. V15 concerns secondary failure propagation: after the secondary process is terminated, the original primary remains green and reports a valid state. The corrected local software path clears validity and drives both rings to all-red.

\begin{figure*}[!t]
  \centering
  \includegraphics[width=\textwidth]{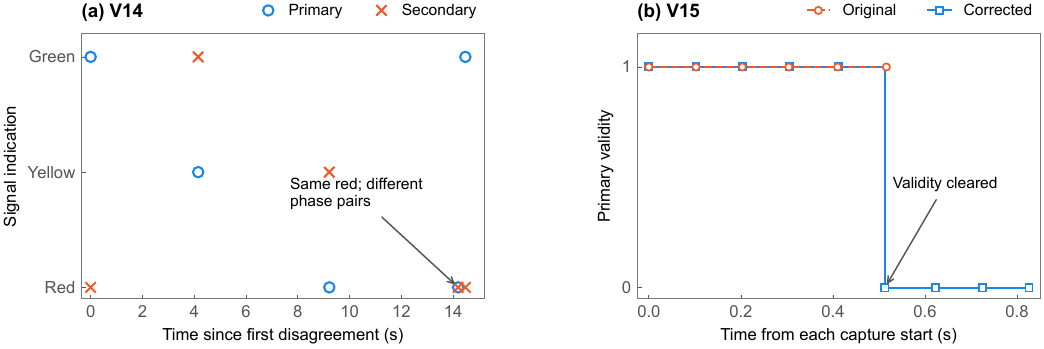}
  \caption{Recorded SCOPE behavior in the time-curve format. (a) V14: primary and secondary indication values at five blind-campaign seed~0 disagreement snapshots; markers show the recorded values. The late red/red snapshot has different phase pairs. The independent actuated baseline has disagreement in all 83 matched samples with primary validity~1. (b) V15: original and corrected primary-validity curves from separate secondary-process-loss captures; time is relative to each capture's first sample. The corrected capture ends all-red with validity~0.}
  \label{fig:redundancy}

\end{figure*}
Figure~\ref{fig:redundancy} shows the recorded disagreement snapshots and process-loss captures.
\subsubsection{Memory and Object Lifetime}
V16 permits an out-of-range pedestrian phase identifier to index beyond the secondary controller's phase array. V17 retains callback references to a controller object after its lifetime has ended. The respective experiments demonstrate out-of-bounds memory access and invalid callback-object access in the secondary-controller software.

V16 checks only whether the supplied phase value is positive before indexing the phase array. Assignment value 17 produces an invalid write in five isolated ASan runs, using a fixture that keeps the controller object alive. V17 constructs the controller as a temporary object and registers its \key{this} pointer in callbacks before the temporary is destroyed. A normal status-request callback produces a stack-use-after-scope report in the original secondary.
\begin{lstlisting}[style=tv,language=C++,
  caption={SCOPE secondary: the pedestrian-phase assignment checks that the
  supplied phase is positive but not that it is within the 1--8 phase array (V16).
  Value 17 produces the out-of-bounds write reported by ASan.},
  label={lst:v16}]
if (value > 0)
{
    // ...
    atc->detector[id].phase = value;
    atc->phaseArray[value - 1]->pedestrianCall =
        &atc->detector[id].call;
}
\end{lstlisting}

\section{TrafficVET Design and Evaluation}
\label{sec:tool}\label{sec:evaluation}
\subsection{TrafficVET Architecture}
\label{sec:flows}
\textbf{Blind campaigns.} The generic generator, generic-v1, reads only the extracted model: signal groups, phases, detector mappings, timing-parameter capabilities, and lifecycle actions. The generator uses no finding catalogue or recorded triggering sequence. Finding matching occurs after execution. The generator, seeds 0--19, and per-surface horizons are frozen in a manifest that records the generator's hash and these invariants. Three findings in Table~\ref{tab:findings} (V04, V05, V13) were first flagged by these campaigns; their mechanisms were isolated afterwards by source analysis and targeted replay (Sections~\ref{sec:v05}, \ref{sec:v13}, \ref{sec:v04}).

Figure~\ref{fig:trafficvet} connects controller artifacts, the extracted safety model, test generation, original-controller execution, and independent safety oracles.
% Retain Figure 5 numbering while placing the architecture beside its first reference.
\setcounter{figure}{4}
\begin{figure*}[!t]
  \centering
  \includegraphics[width=5.2in]{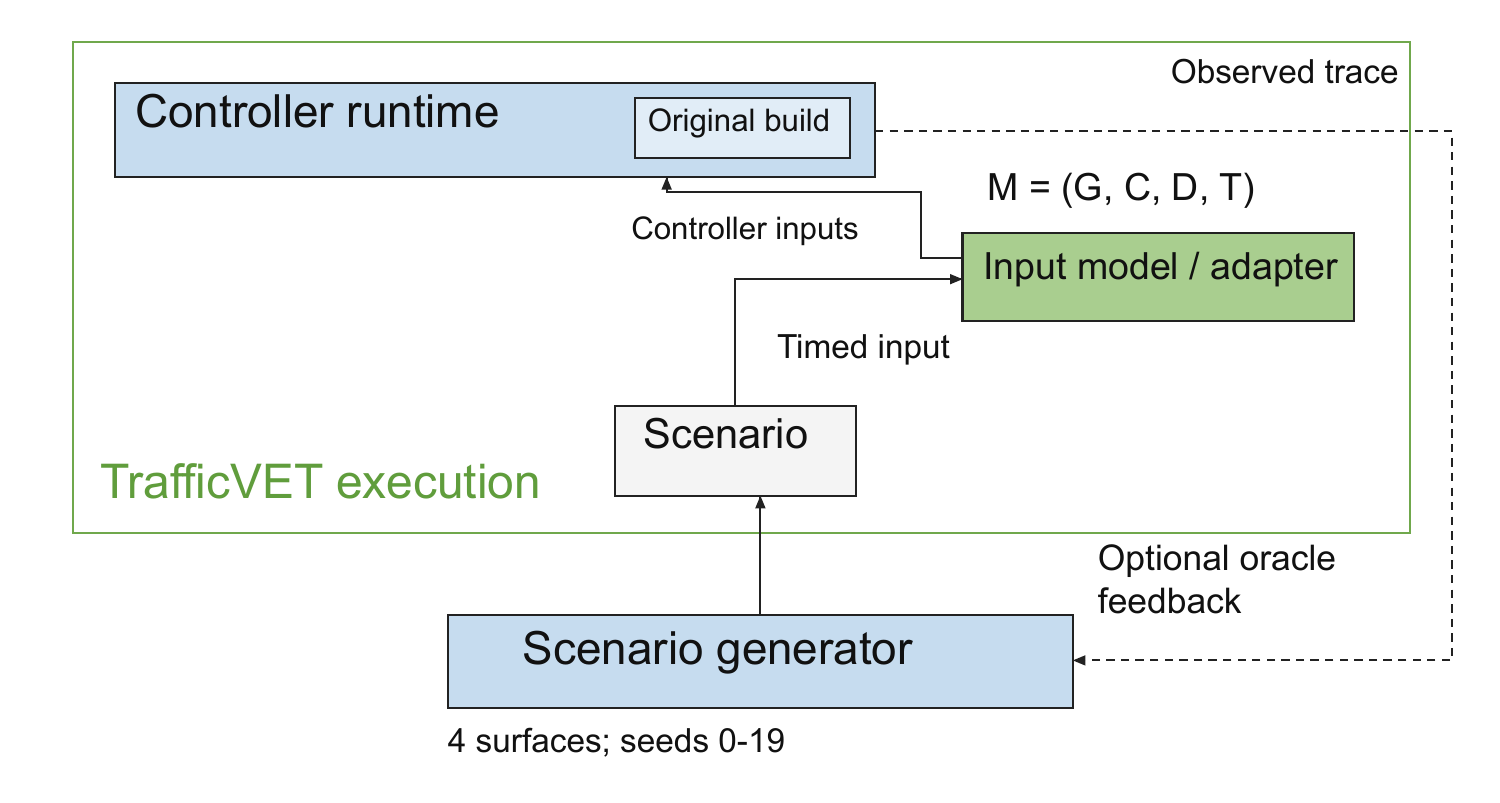}
  \caption{TrafficVET's input-generation and execution loop. A scenario generator uses the extracted model $M=(G,C,D,T)$; input models and target adapters deliver scenarios to original controller builds. Independent safety oracles check observed traces, with optional search feedback shown by the dashed return path.}
  \label{fig:trafficvet}

\end{figure*}
\setcounter{figure}{3}

\argus{} reads source and configuration artifacts, produces a controller model, generates scenarios, executes a target adapter, and evaluates the resulting traces. 

The common model contains signal groups, phase membership, conflict pairs, ordered intergreen values, timing parameters, detector mappings, and target capabilities. A scenario associates input intervals or timed actions with that model. The execution adapter translates a scenario into the input vocabulary of its controller.

\textbf{Constraint Extraction and Scenario Generation.}
For \oc{}, extraction reads the group list, phases, intergreen matrix, and per-group timing fields. Nonzero intergreen entries support a symmetric relation used for configured conflict checking, while the original direction remains in the clearance model. The model also records references to configuration keys in executable source.

For openTLC, extraction reads the conflict list, directional clearances, and detector-to-group mapping. Geometry-derived relationships are available to the oracle. The default generator uses controller inputs and mappings.

For SCOPE, the model captures phase and color symbols, engineering parameters, and the observations available from the primary and secondary traces. The generic generator samples vehicle yellow, red clearance, and pedestrian clearance from the values $\{0,1,2,5,8,10\}$. Pedestrian walk values are sampled from $\{3,5,8,10\}$. Detector channels and pedestrian calls are selected through the target adapter. Sampling zero-valued timing parameters and lifecycle actions enabled the generic campaigns to detect V04 and V13.

 \oc{} scenarios include request pressure, lifecycle actions, and malformed detector inputs; openTLC scenarios generate detector request spans; SCOPE scenarios combine calls and engineering settings. Target capabilities determine which actions are meaningful.

\textbf{Original-Software Execution.}
The toolkit offers fast and real execution modes. The fast \oc{} path drives the original controller core through the existing adapter. The real path starts original Clockwork and NATS, publishes detector and lifecycle inputs, captures group-control messages, and preserves process logs and exit status.

The SCOPE real-mode adapter builds and launches the original 32-bit primary and secondary, drives the original GUI protocol, and collects fresh CSV traces. Each run resets the target by stopping the controller processes before the next launch. The oracle aligns primary and secondary samples by timestamp.

An execution record identifies its input seed, target surface, configuration, trace, and findings. Runtime checks record whether the process exited and the associated output state.

\textbf{Independent Safety Oracles.}
The oracle layer evaluates observed traces outside the target's own decision procedure. For conflicting movements, the property is
\begin{equation}
 C_{ij}=1\ \Longrightarrow\ \neg\bigl(G_i(t)\land G_j(t)\bigr)
\label{eq:conflict}
\end{equation}
where $G_i(t)$ indicates green for movement $i$. The configured relation and the independently examined geometric relation are retained as distinct evidence sources. The resulting finding names the pair whose indications violate the selected relation.

For an ordered intergreen transition, let $e_i$ be the relevant end of green for movement $i$ and $s_j$ a new start of green for movement $j$. A timing check compares
\begin{equation}
 s_j-e_i+\epsilon\ \geq\ D_{ij}
\label{eq:clearance}
\end{equation}
with an adapter-appropriate sampling tolerance $\epsilon$. The Open Controller oracle evaluates the relation when the new green begins. This event-based formulation associates the finding with a transition rather than repeatedly counting each sample of the same active green.

Unsafe-transition checks detect green-to-red changes that omit amber in the observed sequence. Pedestrian checks compare the trace-reported active duration against the supplied walk time plus the profile's 5-second minimum clearance, so a record that supplies a zero clearance is checked against the interval it should have produced rather than the one it requested. Runtime checks record abnormal exit. Controller-lock checks use recurring locked-state observations with a retained next-phase index; the V12 root-evidence check additionally requires the recorded current/next relationship.

Liveness is evaluated per request episode. A request that was served earlier can begin a new unserved episode later. The Open Controller liveness oracle combines a wait threshold based on the configured service cycle with the observed locked or transition state. Adjacent observations of the same event are aggregated with first and last timestamps and a sample count.

\subsection{Test Campaigns}
\label{sec:artifact}
The generic protocol covers four execution surfaces: Open Controller main/fast, Open Controller extender/real, openTLC/fast, and SCOPE/real. Each uses seeds 0--19. Per-seed horizons are 20 seconds, 12 seconds, 2,500 native steps, and 20 seconds, respectively. Campaign inputs and horizons are retained for reproduction~\cite{klees,prudent}. The recorded environment uses Python 3.13.3; SCOPE uses the original 32-bit primary/secondary build, and Open Controller real mode uses original Clockwork with NATS. The campaign stores traces, process exits, and findings before post-campaign evaluation.

Table~\ref{tab:evaluation} summarizes discovery, mechanism reproduction, and targeted confirmation for eight findings. V01 is tested separately through a NATS authorization probe.

 V13 uses ten original and ten fixed runs with emitted group-state messages. V03 uses ten original runs, three 5-second-clearance controls, and a fixed replay. V04 uses ten original and three fixed runs and reads the combined pedestrian-duration field in the runtime trace.

\subsection{TrafficVET Detection Results}
\begin{figure*}[!t]
  \centering
  \includegraphics[width=\textwidth]{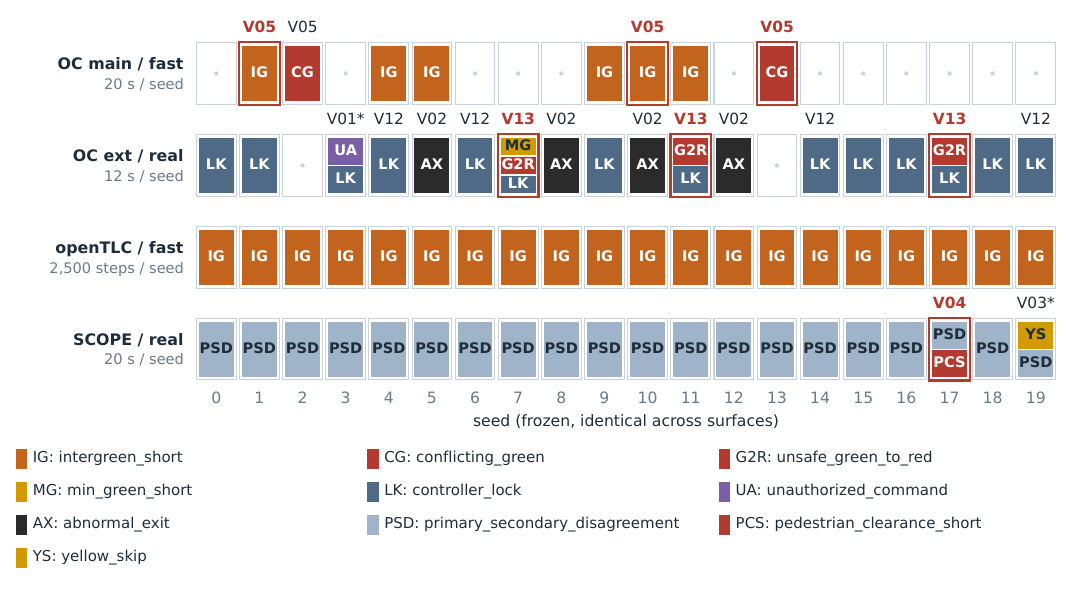}
  \caption{Outcome of the frozen blind campaigns: four surfaces, seeds 0--19,
  one cell per run. Cell colors and abbreviations give the oracle signatures recorded in that
  run; a red frame marks the runs in which a finding was first surfaced
  (V04 at SCOPE seed~17; V13 at extender seeds 7, 11, 17; V05 main-surface
  matches). Labels without a frame mark runs
  that reproduce an already-known root predicate (V02, V12) or only its
  anomaly signature (V01*, V03*). Seed~2 also reproduces the V05 directional conflict predicate. The six
  directional matches comprise two short-clearance records (seeds~1 and~10)
  and four ordered conflict records (two each at seeds~2 and~13).
  Open Controller rows use the corrected oracle campaigns; openTLC and SCOPE
  use their recorded generic campaigns.}
  \label{fig:seedgrid}

\end{figure*}

Figure~\ref{fig:seedgrid} shows oracle signatures across the four campaign surfaces.
Table~\ref{tab:evaluation} summarizes the campaign outcomes. V04, V05, and V13 were first surfaced by blind campaigns and confirmed afterwards as described in Section~\ref{sec:assessment}. V02 is reproduced by four abnormal exits (seeds 5, 8, 10, 12) on a non-numeric detector count; V12 by a persistent locked state with the retained next phase equal to the current phase (seeds 4, 6, 7, 14, 19). The blind campaign produced one yellow-skip signature at seed 19 with a 1.0-second yellow setting. A separate targeted replay established the zero-clearance parameter-validation failure. V11 was evaluated through a 600-second targeted replay, beyond the 20-second generic campaign horizon. V01's root predicate is established by the separate NATS authorization probe; the campaign recorded the unauthorized start at seed 3.

\begin{table*}[!t]
\caption{Findings Surfaced or Reproduced by Blind Campaigns}
\label{tab:evaluation}
\centering\footnotesize
\setlength{\tabcolsep}{3pt}
\renewcommand{\arraystretch}{1.20}
\begin{tabularx}{\textwidth}{@{}L{10mm}L{73mm}L{20mm}Y@{}}
\toprule
Finding & First surfaced by campaign & Root predicate & Targeted confirmation \\
\midrule
V04 & Yes (seed 17) & Yes & 10/10 original, 3/3 fixed \\
V05 & Yes (seeds 1, 10, 13; 6 directional matches) & Yes & original/fixed replay \\
V13 & Yes (seeds 7, 11, 17) & Yes & 10/10 original, 10/10 fixed \\
V02 & --- (reproduced: seeds 5, 8, 10, 12) & Yes & original/fixed replay \\
V12 & --- (reproduced: seeds 4, 6, 7, 14, 19) & Yes & original/fixed trace \\
V03 & --- (signature only, seed 19) & --- & 10/10 original, fixed replay \\
V01 & --- (signature, seed 3) & ACL probe & ACL replay \\
V11 & --- (20 s horizon) & --- & targeted replay \\
\bottomrule
\end{tabularx}
\end{table*}

\section{Cross-Controller Security Implications}
\subsection{Shared Root Causes Across Different Architectures}
Open Controller and openTLC both store directional clearances, but V05 applies the wrong index direction while V07 omits the configured interval from execution. V03 accepts zero vehicle-clearance values before phase logic uses them. In all three controllers, a configured or expected safety rule fails at a concrete data-to-state operation. The corrections act on the index, transition gate, and input assignment, respectively.

\subsection{Distinct Failures with Shared Safety Consequences}
V13 and V03 both produce green-to-red transitions without yellow. V13 reconstructs live state during restart; V03 removes timing through an accepted engineering value. V11 delays a conflicting phase under persistent detector demand. The runtime trace identifies the effect; the exercised input and source operation identify the root cause.

\subsection{Redundant-Controller Security}
SCOPE separates control between Ada and C++, yet its comparison path accepts a wrong-ring field in V09. V14 affects synchronization, and V15 fails to propagate process loss. V16 and V17 affect the secondary's memory integrity. Redundancy therefore depends on the correctness of comparison fields, event ordering, failure latches, and the callback objects that produce the compared state.

\textbf{Memory-Safety Failures and Fail-Safe Propagation.} V16 and V17 identify two independent memory-safety defects in the SCOPE secondary controller. In isolated tests, V16 causes an out-of-bounds memory access and secondary-controller crash, while V17 produces a stack-use-after-scope error during a normal status callback. Both defects were reproduced in five runs. Separately, V15 demonstrates a failure of the original controller's fail-safe mechanism: after the secondary process is terminated, the primary continues displaying green with its validity flag set to 1 instead of entering a safe state.  

\textbf{Timing Manipulation and Startup Validation Weaknesses.} V03 and V04 demonstrate that unauthenticated TCP engineering inputs can modify safety-critical timing parameters, causing skipped vehicle yellow intervals and shortened pedestrian intervals. Independently, V14 demonstrates a startup synchronization and validation failure in the original SCOPE controller. In the actuated baseline, all 83 matched primary-secondary samples show state disagreement, while the primary continues reporting validity as 1. These results establish that unsafe timing parameters are accepted and that the redundant controller can report a valid state despite startup inconsistencies. 

\section{Root-Cause Corrections and Regression Validation}
\label{sec:hardening}
\subsection{Access-Control Enforcement}
V01 separates controller and detector-publisher NATS permissions: restricted stop and output publications are rejected while authorized control remains available~\cite{natsauth,nats,natsmonitor,nist207}. TLS protects transport~\cite{natstls,rfc8446}.
\subsection{Engineering-Input Validation}
V02 validates detector values before conversion and preserves execution on invalid input. V03 replaces nonpositive yellow/red-clearance values with the 5-second profile, restoring yellow with zero replay violations. V04 validates every phase's walk and clearance, restoring 5.0 to 10.0 seconds in all three fixed runs.
\subsection{Signal Timing and Conflict Enforcement}
V05 corrects intergreen index order; V06 moves the timer reference to the signal-state boundary. V07 gates green on ordered clearance and resets timestamp history, yielding zero violations over 20,000 steps. V08 adds the verified crossing relationships and eliminates all three overlaps.
\subsection{Lifecycle and Fail-Safe State Correction}
V10 checks requesting startup groups, removing rapid transitions and conflicting greens. V11 restores phase progression, serving the competing group at 27.83 seconds; V12 clears the stale next-phase index. V13 retains green in all ten corrected runs. V09 corrects the second-ring comparison. V14's injected inconsistent-state correction drives both controllers all-red and clears validity. V15 latches communication failure; its replay ends all-red with validity zero.
\subsection{Memory and Object-Lifetime Correction}
V16 checks phase values against 1--8 before indexing; its isolated corrected run avoids the invalid-write crash. V17 preserves controller lifetime for callback threads and removes the stack-use-after-scope report.

\section{Discussion}
The 17 core findings differ in attacker reachability and experimental scope. Four (V01, V03, V04, and V13) are triggered through unauthenticated network inputs in the tested configurations, while four (V02, V05, V11, and V12) admit detector-publisher input. V16 has an identified local engineering-input path and an isolated out-of-bounds reproduction. Seven findings arise during ordinary controller operation or process failure, and V09 has no established externally controlled trigger. Fourteen findings were exercised at controller level and three (V09, V16, and V17) were validated in isolated component tests.  Open Controller and SCOPE have documented field-test histories.

TrafficVET exposes controller extractors, input generators, adapters, trace schemas, oracles, and evaluation predicates as separate modules. Applying the framework to another controller involves extracting its engineering profile, mapping its input interface, and translating its outputs into signal-state and timing observations. The recorded original-versus-corrected scenarios provide regression inputs for the confirmed failure mechanisms and their safety oracles.

Our coordinated-disclosure plan follows CERT guidance~\cite{certcvd}: share the affected input paths, reproduction evidence, and patches with maintainers, and coordinate publication of actionable details with remediation. For Open Controller, disclosure includes the City of Helsinki and Conveqs as operators of the J{\"a}tk{\"a}saari trial environment, in addition to the repository maintainers.

\section{Conclusion}
We presented a systematic security analysis of three traffic signal controllers and characterized 17 security-relevant software findings.  The findings span five root-cause families, with attacker-access classes and validation scopes recorded alongside CWE-based root causes, triggering conditions, and observed effects.  Our experiments reveal failures in authorization, input handling, signal timing, state execution, redundancy protection, and memory safety. We validate the findings through original software execution, component-level analysis, and corrective experiments, demonstrating effects ranging from conflicting signal indications and reduced clearance intervals to controller unavailability and fail-safe validation failures.

We developed TrafficVET, which extracts controller constraints, drives original controller software, and checks traces with independent oracles. TrafficVET's blind campaigns first identified V04, V05, and V13; subsequent source analysis and targeted replay confirmed their mechanisms. 

\label{sec:bodyend}

\bibliographystyle{IEEEtran}
\bibliography{references}

% Generated by IEEEtran.bst, version: 1.14 (2015/08/26)
\begin{thebibliography}{10}
\providecommand{\url}[1]{#1}
\csname url@samestyle\endcsname
\providecommand{\newblock}{\relax}
\providecommand{\bibinfo}[2]{#2}
\providecommand{\BIBentrySTDinterwordspacing}{\spaceskip=0pt\relax}
\providecommand{\BIBentryALTinterwordstretchfactor}{4}
\providecommand{\BIBentryALTinterwordspacing}{\spaceskip=\fontdimen2\font plus
\BIBentryALTinterwordstretchfactor\fontdimen3\font minus
  \fontdimen4\font\relax}
\providecommand{\BIBforeignlanguage}[2]{{%
\expandafter\ifx\csname l@#1\endcsname\relax
\typeout{** WARNING: IEEEtran.bst: No hyphenation pattern has been}%
\typeout{** loaded for the language `#1'. Using the pattern for}%
\typeout{** the default language instead.}%
\else
\language=\csname l@#1\endcsname
\fi
#2}}
\providecommand{\BIBdecl}{\relax}
\BIBdecl

\bibitem{fhwa}
\BIBentryALTinterwordspacing
{Federal Highway Administration}, ``Traffic signal timing manual: Chapter 5,
  basic signal timing,'' FHWA-HOP-08-024, 2008. [Online]. Available:
  \url{https://ops.fhwa.dot.gov/publications/fhwahop08024/chapter5.htm}
\BIBentrySTDinterwordspacing

\bibitem{yellow}
\BIBentryALTinterwordspacing
------, ``Yellow change intervals,'' Proven Safety Countermeasures, accessed
  October 30, 2026. [Online]. Available:
  \url{https://highways.dot.gov/safety/proven-safety-countermeasures/yellow-change-intervals}
\BIBentrySTDinterwordspacing

\bibitem{mutcd}
\BIBentryALTinterwordspacing
------, ``{Manual on Uniform Traffic Control Devices, 11th Edition with
  Revision 1},'' accessed October 30, 2026. [Online]. Available:
  \url{https://mutcd.fhwa.dot.gov/kno_11th_Editionr1.htm}
\BIBentrySTDinterwordspacing

\bibitem{smartjunction}
\BIBentryALTinterwordspacing
{Aalto University}, ``Smart junction,'' Department of Built Environment.
  [Online]. Available:
  \url{https://www.aalto.fi/en/department-of-built-environment/smart-junction}
\BIBentrySTDinterwordspacing

\bibitem{conveqsjunction}
\BIBentryALTinterwordspacing
{Conveqs Oy}, ``Jätkäsaari smart junction.'' [Online]. Available:
  \url{https://www.conveqs.fi/jatkasaari}
\BIBentrySTDinterwordspacing

\bibitem{conveqstrials2025}
\BIBentryALTinterwordspacing
K.~Koskinen, ``Aalto-yhteisty{"o}: Verkottuneet autot ({V2X}),'' Conveqs
  presentation, TECHBOOST day, April 3, 2025. [Online]. Available:
  \url{https://www.techboost.fi/wp-content/uploads/2025/04/Conveqskalvot-TECHBOOST-pa%CC%88iva%CC%88.pdf}
\BIBentrySTDinterwordspacing

\bibitem{opencontroller}
\BIBentryALTinterwordspacing
{Open-TLC}, ``Open controller: Clockwork traffic controller,'' Project
  repository, accessed October 30, 2026. [Online]. Available:
  \url{https://github.com/Open-TLC/open_controller}
\BIBentrySTDinterwordspacing

\bibitem{scopefield}
\BIBentryALTinterwordspacing
{Advanced Technologies, Incorporated}, ``Signal state transition software
  development: {NTCIP} interface and live intersection testing,'' Advanced
  Technologies, Incorporated, Phase II B Final Report SBIR 06-FH1, Feb. 2013.
  [Online]. Available:
  \url{https://www.advancedtechcorp.com/SCOPE/assets/files/SBIR_Final_Report_Phase_II_B.pdf}
\BIBentrySTDinterwordspacing

\bibitem{icsdefense}
\BIBentryALTinterwordspacing
{Cybersecurity and Infrastructure Security Agency}, ``{Improving Security of
  Open Source Software in Operational Technology and Industrial Control
  Systems},'' accessed October 30, 2026. [Online]. Available:
  \url{https://www.cisa.gov/sites/default/files/2023-10/Fact%20Sheet%20-%20Improving%20OSS%20in%20OT%20ICS%20-%20508c.pdf}
\BIBentrySTDinterwordspacing

\bibitem{nist218}
\BIBentryALTinterwordspacing
{National Institute of Standards and Technology}, ``{Secure Software
  Development Framework (SSDF) Version 1.1},'' accessed October 30, 2026.
  [Online]. Available: \url{https://csrc.nist.gov/pubs/sp/800/218/final}
\BIBentrySTDinterwordspacing

\bibitem{opentlc}
\BIBentryALTinterwordspacing
M.~Harmenzon, ``{openTLC}: An open source traffic light control program,''
  Project repository, accessed October 30, 2026. [Online]. Available:
  \url{https://github.com/MartijnHarmenzon/openTLC}
\BIBentrySTDinterwordspacing

\bibitem{scope}
\BIBentryALTinterwordspacing
{Advanced Technologies, Incorporated}, ``{ATI\_SCOPE}: Open source traffic
  control software,'' SourceForge project, accessed October 30, 2026. [Online].
  Available: \url{https://sourceforge.net/projects/atiscope/}
\BIBentrySTDinterwordspacing

\bibitem{ghena2014}
\BIBentryALTinterwordspacing
B.~Ghena, W.~Beyer, A.~Hillaker, J.~Pevarnek, and J.~A. Halderman, ``Green
  lights forever: Analyzing the security of traffic infrastructure,'' in
  \emph{8th USENIX Workshop on Offensive Technologies (WOOT)}, 2014. [Online].
  Available:
  \url{https://www.usenix.org/conference/woot14/workshop-program/presentation/ghena}
\BIBentrySTDinterwordspacing

\bibitem{chen2018}
Q.~A. Chen, Y.~Yin, Y.~Feng, Z.~M. Mao, and H.~X. Liu, ``Exposing congestion
  attack on emerging connected vehicle based traffic signal control,'' in
  \emph{Network and Distributed System Security Symposium (NDSS)}, 2018.

\bibitem{li2025}
\BIBentryALTinterwordspacing
W.~Li, T.~Zhu, and Y.~Feng, ``{WIP}: Learning adversarial attacks on adaptive
  traffic signal control systems under cooperative perception,'' in \emph{3rd
  USENIX Symposium on Vehicle Security and Privacy (VehicleSec)}, 2025, pp.
  343--351. [Online]. Available:
  \url{https://www.usenix.org/conference/vehiclesec25/presentation/li-wangzhi}
\BIBentrySTDinterwordspacing

\bibitem{signalhandbook}
\BIBentryALTinterwordspacing
{Federal Highway Administration}, ``{Traffic Control Systems Handbook},''
  accessed October 30, 2026. [Online]. Available:
  \url{https://rosap.ntl.bts.gov/view/dot/16624}
\BIBentrySTDinterwordspacing

\bibitem{detectorhandbook}
\BIBentryALTinterwordspacing
------, ``{Traffic Detector Handbook: Third Edition},'' accessed October 30,
  2026. [Online]. Available:
  \url{https://www.fhwa.dot.gov/publications/research/operations/its/06108/}
\BIBentrySTDinterwordspacing

\bibitem{lattu2025}
\BIBentryALTinterwordspacing
O.~Lattu, ``Machine learning-based optimization model in traffic signal
  control,'' Bachelor's thesis, Aalto University, 2025, in Finnish; Finnish
  title: Koneoppimiseen perustuva optimointimalli liikenteen valo-ohjauksessa.
  [Online]. Available:
  \url{https://www.finna.fi/Record/aaltodoc.123456789_140234}
\BIBentrySTDinterwordspacing

\bibitem{sumo}
\BIBentryALTinterwordspacing
{Eclipse Foundation}, ``About eclipse {SUMO},'' accessed October 30, 2026.
  [Online]. Available: \url{https://eclipse.dev/sumo/about/}
\BIBentrySTDinterwordspacing

\bibitem{scopeprogram}
\BIBentryALTinterwordspacing
{Advanced Technologies, Incorporated}, ``Traffic intersection control logic,''
  accessed October 30, 2026. [Online]. Available:
  \url{https://www.advancedtechcorp.com/page2.html}
\BIBentrySTDinterwordspacing

\bibitem{natssubject}
\BIBentryALTinterwordspacing
{NATS Authors}, ``{Subject-Based Messaging},'' accessed October 30, 2026.
  [Online]. Available: \url{https://docs.nats.io/nats-concepts/subjects}
\BIBentrySTDinterwordspacing

\bibitem{natsprotocol}
\BIBentryALTinterwordspacing
------, ``{Client Protocol},'' accessed October 30, 2026. [Online]. Available:
  \url{https://docs.nats.io/reference/reference-protocols/nats-protocol}
\BIBentrySTDinterwordspacing

\bibitem{rfc9293}
\BIBentryALTinterwordspacing
{Wesley Eddy}, ``{Transmission Control Protocol (TCP)},'' accessed October 30,
  2026. [Online]. Available: \url{https://www.rfc-editor.org/rfc/rfc9293.html}
\BIBentrySTDinterwordspacing

\bibitem{rfc8259}
\BIBentryALTinterwordspacing
{Tim Bray}, ``{The JavaScript Object Notation (JSON) Data Interchange
  Format},'' accessed October 30, 2026. [Online]. Available:
  \url{https://www.rfc-editor.org/rfc/rfc8259.html}
\BIBentrySTDinterwordspacing

\bibitem{yaml}
\BIBentryALTinterwordspacing
{YAML Language Development Team}, ``{YAML Ain't Markup Language (YAML) Version
  1.2.2},'' accessed October 30, 2026. [Online]. Available:
  \url{https://yaml.org/spec/1.2.2/}
\BIBentrySTDinterwordspacing

\bibitem{sumonet}
\BIBentryALTinterwordspacing
{Eclipse SUMO Authors}, ``{SUMO Road Networks},'' accessed October 30, 2026.
  [Online]. Available:
  \url{https://sumo.dlr.de/docs/Networks/SUMO_Road_Networks.html}
\BIBentrySTDinterwordspacing

\bibitem{sumotl}
\BIBentryALTinterwordspacing
------, ``{Traffic Lights},'' accessed October 30, 2026. [Online]. Available:
  \url{https://sumo.dlr.de/docs/Simulation/Traffic_Lights.html}
\BIBentrySTDinterwordspacing

\bibitem{traci}
\BIBentryALTinterwordspacing
------, ``{TraCI},'' accessed October 30, 2026. [Online]. Available:
  \url{https://sumo.dlr.de/docs/TraCI/Interfacing_TraCI_from_Python.html}
\BIBentrySTDinterwordspacing

\bibitem{atspm}
\BIBentryALTinterwordspacing
{Federal Highway Administration}, ``{Automated Traffic Signal Performance
  Measures},'' accessed October 30, 2026. [Online]. Available:
  \url{https://ops.fhwa.dot.gov/arterial_mgmt/performance_measures.htm}
\BIBentrySTDinterwordspacing

\bibitem{signalsafety}
\BIBentryALTinterwordspacing
------, ``{Traffic Signal Program Handbook},'' accessed October 30, 2026.
  [Online]. Available:
  \url{https://ops.fhwa.dot.gov/publications/fhwahop23041/fhwahop23041.pdf}
\BIBentrySTDinterwordspacing

\bibitem{huang2021}
\BIBentryALTinterwordspacing
S.~E. Huang, W.~Wong, Y.~Feng, Q.~A. Chen, Z.~M. Mao, and H.~X. Liu, ``{Impact
  Evaluation of Falsified Data Attacks on Connected Vehicle Based Traffic
  Signal Control},'' arXiv:2010.04753, 2020. [Online]. Available:
  \url{https://arxiv.org/abs/2010.04753}
\BIBentrySTDinterwordspacing

\bibitem{dasgupta2021}
\BIBentryALTinterwordspacing
S.~Dasgupta, C.~Hollis, M.~Rahman, and T.~Atkison, ``{An Innovative Attack
  Modelling and Attack Detection Approach for a Waiting Time-based Adaptive
  Traffic Signal Controller},'' arXiv:2108.08627, 2021. [Online]. Available:
  \url{https://arxiv.org/abs/2108.08627}
\BIBentrySTDinterwordspacing

\bibitem{irfan2022}
\BIBentryALTinterwordspacing
M.~S. Irfan, M.~Rahman, T.~Atkison, S.~Dasgupta, and A.~Hainen,
  ``{Reinforcement Learning based Cyberattack Model for Adaptive Traffic Signal
  Controller in Connected Transportation Systems},'' arXiv:2211.01845, 2022.
  [Online]. Available: \url{https://arxiv.org/abs/2211.01845}
\BIBentrySTDinterwordspacing

\bibitem{thodi2020}
\BIBentryALTinterwordspacing
B.~T. Thodi, T.~Mulumba, and S.~E. Jabari, ``{Noticeability Versus Impact in
  Traffic Signal Tampering},'' arXiv:2002.03119, 2020. [Online]. Available:
  \url{https://arxiv.org/abs/2002.03119}
\BIBentrySTDinterwordspacing

\bibitem{securitygame}
\BIBentryALTinterwordspacing
A.~Laszka, W.~Abbas, Y.~Vorobeychik, and X.~Koutsoukos, ``{Detection and
  Mitigation of Attacks on Transportation Networks as a Multi-Stage Security
  Game},'' arXiv:1808.08349, 2018. [Online]. Available:
  \url{https://arxiv.org/abs/1808.08349}
\BIBentrySTDinterwordspacing

\bibitem{resilience}
\BIBentryALTinterwordspacing
R.~A. Mallah, T.~Halabi, and B.~Farooq, ``{Resilience-by-design in Adaptive
  Multi-Agent Traffic Control Systems},'' arXiv:2012.02675, 2020. [Online].
  Available: \url{https://arxiv.org/abs/2012.02675}
\BIBentrySTDinterwordspacing

\bibitem{laser}
\BIBentryALTinterwordspacing
S.~H.~V. Bhupathiraju, T.~Sato, M.~Clifford, T.~Sugawara, Q.~A. Chen, and
  S.~Rampazzi, ``{On the Vulnerability of Traffic Light Recognition Systems to
  Laser Illumination Attacks},'' in \emph{Symposium on Vehicle Security and
  Privacy}, 2024. [Online]. Available:
  \url{https://www.ndss-symposium.org/ndss-paper/auto-draft-451/}
\BIBentrySTDinterwordspacing

\bibitem{rvfuzzer}
\BIBentryALTinterwordspacing
T.~Kim, C.~H. Kim, J.~Rhee, F.~Fei, Z.~Tu, G.~Walkup, X.~Zhang, X.~Deng, and
  D.~Xu, ``{RVFuzzer}: Finding input validation bugs in robotic vehicles
  through control-guided testing,'' in \emph{28th USENIX Security Symposium},
  2019, pp. 425--442. [Online]. Available:
  \url{https://www.usenix.org/conference/usenixsecurity19/presentation/kim}
\BIBentrySTDinterwordspacing

\bibitem{pgfuzz}
\BIBentryALTinterwordspacing
H.~Kim, M.~O. Ozmen, A.~Bianchi, Z.~B. Celik, and D.~Xu, ``{PGFUZZ}:
  Policy-guided fuzzing for robotic vehicles,'' in \emph{Network and
  Distributed System Security Symposium (NDSS)}, 2021. [Online]. Available:
  \url{https://www.ndss-symposium.org/ndss-paper/pgfuzz-policy-guided-fuzzing-for-robotic-vehicles/}
\BIBentrySTDinterwordspacing

\bibitem{aflgo}
M.~B{\"o}hme, V.-T. Pham, M.-D. Nguyen, and A.~Roychoudhury, ``Directed greybox
  fuzzing,'' in \emph{ACM SIGSAC Conference on Computer and Communications
  Security (CCS)}, 2017, pp. 2329--2344.

\bibitem{savior}
\BIBentryALTinterwordspacing
R.~Quinonez, J.~Giraldo, L.~Salazar, E.~Bauman, A.~Cardenas, and Z.~Lin,
  ``{{SAVIOR}: Securing Autonomous Vehicles with Robust Physical Invariants},''
  in \emph{29th USENIX Security Symposium (USENIX Security 20)}, 2020, pp.
  895--912. [Online]. Available:
  \url{https://www.usenix.org/conference/usenixsecurity20/presentation/quinonez}
\BIBentrySTDinterwordspacing

\bibitem{drivefuzz}
\BIBentryALTinterwordspacing
S.~Kim, M.~Liu, J.~J. Rhee, Y.~Jeon, Y.~Kwon, and C.~H. Kim, ``{DriveFuzz:
  Discovering Autonomous Driving Bugs through Driving Quality-Guided
  Fuzzing},'' arXiv:2211.01829, 2022. [Online]. Available:
  \url{https://arxiv.org/abs/2211.01829}
\BIBentrySTDinterwordspacing

\bibitem{icsref}
\BIBentryALTinterwordspacing
A.~Keliris and M.~Maniatakos, ``{ICSREF: A Framework for Automated Reverse
  Engineering of Industrial Control Systems Binaries},'' in \emph{Network and
  Distributed System Security Symposium}, 2019. [Online]. Available:
  \url{https://www.ndss-symposium.org/ndss-paper/icsref-a-framework-for-automated-reverse-engineering-of-industrial-control-systems-binaries/}
\BIBentrySTDinterwordspacing

\bibitem{icsfuzz}
\BIBentryALTinterwordspacing
D.~Tychalas, H.~Benkraouda, and M.~Maniatakos, ``{{ICSFuzz}: Manipulating
  {I/Os} and Repurposing Binary Code to Enable Instrumented Fuzzing in {ICS}
  Control Applications},'' in \emph{30th USENIX Security Symposium (USENIX
  Security 21)}, 2021, pp. 2847--2862. [Online]. Available:
  \url{https://www.usenix.org/conference/usenixsecurity21/presentation/tychalas}
\BIBentrySTDinterwordspacing

\bibitem{firmadyne}
\BIBentryALTinterwordspacing
D.~D. Chen, M.~Egele, M.~Woo, and D.~Brumley, ``{Towards Automated Dynamic
  Analysis for Linux-based Embedded Firmware},'' in \emph{Network and
  Distributed System Security Symposium}, 2016. [Online]. Available:
  \url{https://www.ndss-symposium.org/wp-content/uploads/2017/09/towards-automated-dynamic-analysis-linux-based-embedded-firmware.pdf}
\BIBentrySTDinterwordspacing

\bibitem{firmafl}
\BIBentryALTinterwordspacing
Y.~Zheng, A.~Davanian, H.~Yin, C.~Song, H.~Zhu, and L.~Sun, ``{{FIRM-AFL}:
  {High-Throughput} Greybox Fuzzing of {IoT} Firmware via Augmented Process
  Emulation},'' in \emph{28th USENIX Security Symposium (USENIX Security 19)},
  2019, pp. 1099--1114. [Online]. Available:
  \url{https://www.usenix.org/conference/usenixsecurity19/presentation/zheng}
\BIBentrySTDinterwordspacing

\bibitem{p2im}
\BIBentryALTinterwordspacing
B.~Feng, A.~Mera, and L.~Lu, ``{{P2IM}: Scalable and Hardware-independent
  Firmware Testing via Automatic Peripheral Interface Modeling},'' in
  \emph{29th USENIX Security Symposium (USENIX Security 20)}, 2020, pp.
  1237--1254. [Online]. Available:
  \url{https://www.usenix.org/conference/usenixsecurity20/presentation/feng}
\BIBentrySTDinterwordspacing

\bibitem{fuzzware}
\BIBentryALTinterwordspacing
T.~Scharnowski, N.~Bars, M.~Schloegel, E.~Gustafson, M.~Muench, G.~Vigna,
  C.~Kruegel, T.~Holz, and A.~Abbasi, ``{Fuzzware: Using Precise {MMIO}
  Modeling for Effective Firmware Fuzzing},'' in \emph{31st USENIX Security
  Symposium (USENIX Security 22)}, 2022, pp. 1239--1256. [Online]. Available:
  \url{https://www.usenix.org/conference/usenixsecurity22/presentation/scharnowski}
\BIBentrySTDinterwordspacing

\bibitem{aflnet}
\BIBentryALTinterwordspacing
R.~Meng, V.-T. Pham, M.~B{\"o}hme, and A.~Roychoudhury, ``{AFLNet Five Years
  Later: On Coverage-Guided Protocol Fuzzing},'' arXiv:2412.20324, 2024.
  [Online]. Available: \url{https://arxiv.org/abs/2412.20324}
\BIBentrySTDinterwordspacing

\bibitem{stateafl}
\BIBentryALTinterwordspacing
R.~Natella, ``{StateAFL: Greybox Fuzzing for Stateful Network Servers},''
  arXiv:2110.06253, 2021. [Online]. Available:
  \url{https://arxiv.org/abs/2110.06253}
\BIBentrySTDinterwordspacing

\bibitem{sgfuzz}
\BIBentryALTinterwordspacing
J.~Ba, M.~B{\"o}hme, Z.~Mirzamomen, and A.~Roychoudhury, ``{Stateful Greybox
  Fuzzing},'' in \emph{31st USENIX Security Symposium (USENIX Security 22)},
  2022, pp. 3255--3272. [Online]. Available:
  \url{https://www.usenix.org/conference/usenixsecurity22/presentation/ba}
\BIBentrySTDinterwordspacing

\bibitem{aflpp}
\BIBentryALTinterwordspacing
{AFL++ Authors}, ``{AFL++ Documentation},'' accessed October 30, 2026.
  [Online]. Available: \url{https://aflplus.plus/docs/}
\BIBentrySTDinterwordspacing

\bibitem{libfuzzer}
\BIBentryALTinterwordspacing
{LLVM Project}, ``{libFuzzer: A Library for Coverage-Guided Fuzz Testing},''
  accessed October 30, 2026. [Online]. Available:
  \url{https://llvm.org/docs/LibFuzzer.html}
\BIBentrySTDinterwordspacing

\bibitem{asanpaper}
\BIBentryALTinterwordspacing
K.~Serebryany, D.~Bruening, A.~Potapenko, and D.~Vyukov, ``{AddressSanitizer: A
  Fast Address Sanity Checker},'' in \emph{USENIX Annual Technical Conference},
  2012. [Online]. Available:
  \url{https://www.usenix.org/conference/atc12/technical-sessions/presentation/serebryany}
\BIBentrySTDinterwordspacing

\bibitem{asan}
\BIBentryALTinterwordspacing
{LLVM Project}, ``{AddressSanitizer},'' accessed October 30, 2026. [Online].
  Available: \url{https://clang.llvm.org/docs/AddressSanitizer.html}
\BIBentrySTDinterwordspacing

\bibitem{cwe862}
\BIBentryALTinterwordspacing
{MITRE}, ``{CWE-862: Missing Authorization},'' accessed October 30, 2026.
  [Online]. Available: \url{https://cwe.mitre.org/data/definitions/862.html}
\BIBentrySTDinterwordspacing

\bibitem{cwe841}
\BIBentryALTinterwordspacing
------, ``{CWE-841: Improper Enforcement of Behavioral Workflow},'' accessed
  October 30, 2026. [Online]. Available:
  \url{https://cwe.mitre.org/data/definitions/841.html}
\BIBentrySTDinterwordspacing

\bibitem{cwe20}
\BIBentryALTinterwordspacing
------, ``{CWE-20: Improper Input Validation},'' accessed October 30, 2026.
  [Online]. Available: \url{https://cwe.mitre.org/data/definitions/20.html}
\BIBentrySTDinterwordspacing

\bibitem{cwe754}
\BIBentryALTinterwordspacing
------, ``{CWE-754: Improper Check for Unusual or Exceptional Conditions},''
  accessed October 30, 2026. [Online]. Available:
  \url{https://cwe.mitre.org/data/definitions/754.html}
\BIBentrySTDinterwordspacing

\bibitem{cwe306}
\BIBentryALTinterwordspacing
------, ``{CWE-306: Missing Authentication for Critical Functionality},''
  accessed October 30, 2026. [Online]. Available:
  \url{https://cwe.mitre.org/data/definitions/306.html}
\BIBentrySTDinterwordspacing

\bibitem{cwe682}
\BIBentryALTinterwordspacing
------, ``{CWE-682: Incorrect Calculation},'' accessed October 30, 2026.
  [Online]. Available: \url{https://cwe.mitre.org/data/definitions/682.html}
\BIBentrySTDinterwordspacing

\bibitem{nats}
\BIBentryALTinterwordspacing
{NATS Authors}, ``Authorization,'' NATS Documentation, accessed October 30,
  2026. [Online]. Available:
  \url{https://docs.nats.io/learn/security/authorization}
\BIBentrySTDinterwordspacing

\bibitem{docker}
\BIBentryALTinterwordspacing
{Docker}, ``{Networking},'' accessed October 30, 2026. [Online]. Available:
  \url{https://docs.docker.com/engine/network/}
\BIBentrySTDinterwordspacing

\bibitem{klees}
\BIBentryALTinterwordspacing
G.~Klees, A.~Ruef, B.~Cooper, S.~Wei, and M.~Hicks, ``{Evaluating Fuzz
  Testing},'' arXiv:1808.09700, 2018. [Online]. Available:
  \url{https://arxiv.org/abs/1808.09700}
\BIBentrySTDinterwordspacing

\bibitem{prudent}
\BIBentryALTinterwordspacing
M.~Schloegel, N.~Bars, N.~Schiller, L.~Bernhard, T.~Scharnowski, A.~Crump,
  A.~A. Ebrahim, N.~Bissantz, M.~Muench, and T.~Holz, ``{SoK: Prudent
  Evaluation Practices for Fuzzing},'' arXiv:2405.10220, 2024. [Online].
  Available: \url{https://arxiv.org/abs/2405.10220}
\BIBentrySTDinterwordspacing

\bibitem{natsauth}
\BIBentryALTinterwordspacing
{NATS Authors}, ``{Authentication},'' accessed October 30, 2026. [Online].
  Available:
  \url{https://docs.nats.io/running-a-nats-service/configuration/securing_nats/auth_intro}
\BIBentrySTDinterwordspacing

\bibitem{natsmonitor}
\BIBentryALTinterwordspacing
------, ``{Monitoring},'' accessed October 30, 2026. [Online]. Available:
  \url{https://docs.nats.io/running-a-nats-service/nats_admin/monitoring}
\BIBentrySTDinterwordspacing

\bibitem{nist207}
\BIBentryALTinterwordspacing
{National Institute of Standards and Technology}, ``{Zero Trust
  Architecture},'' accessed October 30, 2026. [Online]. Available:
  \url{https://csrc.nist.gov/pubs/sp/800/207/final}
\BIBentrySTDinterwordspacing

\bibitem{natstls}
\BIBentryALTinterwordspacing
{NATS Authors}, ``{TLS},'' accessed October 30, 2026. [Online]. Available:
  \url{https://docs.nats.io/running-a-nats-service/configuration/securing_nats/tls}
\BIBentrySTDinterwordspacing

\bibitem{rfc8446}
\BIBentryALTinterwordspacing
{Eric Rescorla}, ``{The Transport Layer Security (TLS) Protocol Version 1.3},''
  accessed October 30, 2026. [Online]. Available:
  \url{https://www.rfc-editor.org/rfc/rfc8446.html}
\BIBentrySTDinterwordspacing

\bibitem{certcvd}
\BIBentryALTinterwordspacing
{Software Engineering Institute}, ``{The CERT Guide to Coordinated
  Vulnerability Disclosure},'' accessed October 30, 2026. [Online]. Available:
  \url{https://www.sei.cmu.edu/library/the-cert-guide-to-coordinated-vulnerability-disclosure/}
\BIBentrySTDinterwordspacing

\end{thebibliography}

\end{document}